\documentclass[12pt,preprint]{aastex}

\begin{document}
\def\lsun{{\rm L_{\odot}}}
\def\msun{{\rm M_{\odot}}}
\def\rsun{{\rm R_{\odot}}}
\def\go{
\mathrel{\raise.3ex\hbox{$>$}\mkern-14mu\lower0.6ex\hbox{$\sim$}}
}
\def\lo{
\mathrel{\raise.3ex\hbox{$<$}\mkern-14mu\lower0.6ex\hbox{$\sim$}}
}
\def\simeq{
\mathrel{\raise.3ex\hbox{$\sim$}\mkern-14mu\lower0.4ex\hbox{$-$}}
}

\input epsf.sty

\title{Gamma-ray bursts, supernova kicks, and
gravitational radiation} \author{Melvyn B.\
Davies, Andrew King, Stephan Rosswog, and Graham
Wynn\\ Department Physics \& Astronomy, University of
Leicester, Leicester LE1 7RH, UK}


\begin{abstract}
We suggest that the collapsing core of a massive rotating star may
fragment to produce two or more compact objects. Their coalescence
under gravitational radiation gives the resulting black hole or
neutron star a significant kick velocity, which may explain those
observed in pulsars. A gamma--ray burst can result only when this kick
is small. Thus only a small fraction of core--collapse supernovae
produce gamma--ray bursts. The burst may be delayed significantly
(hours -- days) after the supernova, as suggested by recent
observations. If this picture is correct, 
core--collapse supernovae
should be significant sources of gravitational radiation with a chirp
signal similar to a coalescing binary.
\end{abstract}

\section{Introduction}

It is now widely believed that long ($\ga $5 s)
gamma--ray bursts are produced by a class of supernovae, known as 
collapsars or hypernovae
(Woosley 1993; MacFadyen \& Woosley, 1999; Paczy\'nski, 1998).
The collapse of the core of
a massive star is assumed to lead to the formation of a black hole, 
the remaining core material having enough angular momentum to
form a massive accreting torus around it. The gravitational energy of
the torus is radiated as neutrinos or converted to a beamed outflow by
MHD processes. An evacuated channel forms along the rotation axis of
the core, allowing the expulsion of matter with high Lorentz factors.

Direct evidence for the association of SNe and GRBs comes from 
the detection of bumps in the afterglow of several gamma--ray bursts 
(e.g. Price et al. 2002 and references therein) and
the recent detection of SN ejecta in the X--ray afterglow of GRB 011211 by
Reeves et al. (2002). However the hypernova class is extremely small:
even allowing for the probable beaming of gamma--ray bursts (Frail et al.
2001), the fraction of HNe among SNe cannot be greater than about
$10^{-3}$. Evidently the production of a gamma--ray burst by a
supernova is a very rare event. What causes this rarity is unclear.

The X--ray observation of a SN--GRB association by Reeves et
al. (2002) throws up a further puzzle. Light travel arguments give a
size $10^{14} - 10^{15}$~cm for the reprocessing region producing the
X--ray spectrum, depending on the beaming. This is much larger than
the radius of the progenitor star,
and must be associated with the
supernova outflow. Indeed the measured blueshift of the spectrum with
respect to the known GRB redshift implies an outflow velocity $\sim
0.1c$. But these two measurements together require that the GRB
occurred between 10~hr and 4~d after the supernova. This is clearly
incompatible with the simplest version of how a hypernova proceeds.

In this paper we offer a solution to both problems. We reconsider the
collapse of a rotating core and suggest by analogy with simulations of
star formation that this may produce two or more compact objects. The
subsequent coalescence of these objects can power a gamma--ray burst,
accounting for the SN--GRB delay. The merger itself will generally
give the black hole resulting from the collapse a significant velocity
(`kick'). This may be the explanation for the kicks observed
in pulsars (Arzoumanian, Chernoff \& Cordes, 2002). 
Following the suggestion of MacFadyen and Woosley (1999) that
GRB production will be adversely affected by such kicks, we show that
only a small fraction of core--collapse
supernovae will produce gamma--ray bursts. These are likely to be
a subset of those producing a  massive black hole 
($\go$ 12 M$_\odot$). 

\section{Core collapse and fragmentation}

It is well known that dynamical collapse of a self--gravitating gas
cloud increases the importance of rotation. The ratio of kinetic to
gravitational binding energy grows as $\sim 1/r$, where $r$ is the
lengthscale of the collapsing object. Many authors (see Bonnell \&
Pringle, 1995 and references therein) have suggested that this probably
leads to fragmentation, seen for example in the collapse of molecular
clouds to form pre--main--sequence stars. Fragmentation requires that
the collapsing core becomes bar--unstable, and that any bar lives a
few dynamical times. In core collapse to nuclear densities the second
requirement is very likely to be met (Bonnell \& Pringle, 1995) while
the first depends on the equation of state and the initial
conditions. Thus determining the precise conditions under which
fragmentation occurs requires large--scale numerical simulations,
which are under way. For the remainder of this paper we consider the
case where two compact objects form with masses and radii $M_1, M_2$
and $R_1, R_2$, with $M_1 > M_2$. The minimum mass of these objects
is set by 
the requirement that  a nucleon fluid exists in their cores.
In Fig 1 we plot $R_2$ as a function of $M_2$ for the equation of state (EOS)
of Shen et al. (1998a,b).  
From Figure 1 we see that we require a
mass $\ga 0.2$ M$_\odot$ -- at lower masses nuclei form 
even in the center of the stars
and the object has a much larger radius. 
A similar low--mass limit is obtained for other equations of state.

\begin{figure}
\plotone{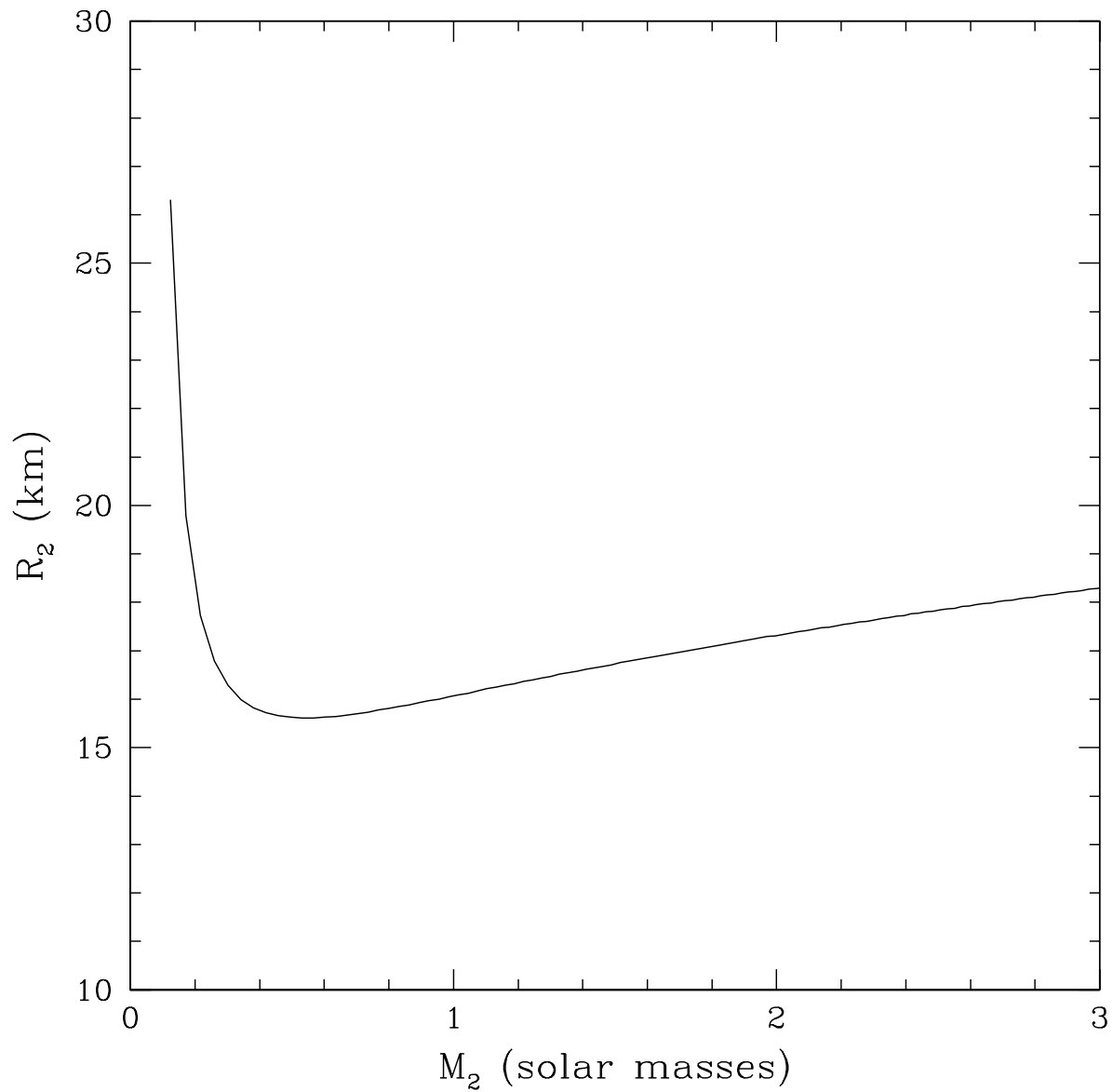}
\caption{Mass--radius relation for cold neutron stars in beta-equilibrium
using the EOS of Shen et al. (1998). The end point on the low mass side 
is reached once nuclei start to form in the center of the star.} 
\end{figure}

\section{Mergers and gamma--ray bursts}

To have the potential of powering a gamma--ray burst, the merger of
the two orbiting lumps must produce a central object surrounded by a
torus. This will happen if at least one of the lumps is a neutron
star rather than a black hole, and mass transfer eventually becomes
dynamically unstable. For a corotating object less massive than the
accretor and filling its Roche lobe, of radius (Paczy\'nski 1971)
\begin{equation}
R_{\rm L} = 0.462\biggl({M_2\over M_1 + M_2}\biggr)^{1/3}a
\label{rl} 
\end{equation} 
(where $a$ is the separation) this requires that $R_{\rm L}$ moves
inwards with respect to its radius $R_2$. The standard result
(e.g. van Teeseling \& King, 1998) is
\begin{equation}
{\dot R_{\rm L}\over R_{\rm L}} - {\dot R_2\over R_2} 
= -{2\dot M_2\over M_2}\biggl({5\over 6} + {\zeta\over 2} - 
{M_2\over M_1}\biggr) + 2{\dot J\over J}, \label{rldot}
\end{equation}
where the mass--radius relation is taken as $R_2 \propto M_2^{\zeta}$,
and $\dot J$ includes all forms of orbital angular momentum loss. This
expression shows that dynamical instability must occur if $\zeta <
-5/3$, since $\dot M_2, \dot J < 0$. The mass--radius relation Fig. 1
now shows that the instability is inevitable for any lobe--filling
object, since it will occur at the latest once its mass is reduced to
$M_2 \simeq 0.2$M$_{\odot}$ and the lump begins to expand rapidly on
mass loss. The tidal lobe of a non--corotating object
is similar in size, so again instability
will occur for $M_2 \simeq 0.2$M$_{\odot}$.
Instability may well happen before this point for 
other reasons:
for example, the orbit may be so close that the accreting
matter cannot form a disc around the accretor, adding a
dynamical--timescale term to $\dot J$. 
For sufficiently stiff equations of state Newtonian
tidal effects can also lead to an instability on a dynamical time scale 
(Lai, Rasio and Shapiro 1993).

The only way that dynamical instability can be avoided is if (a) both
lumps are already black holes, or (b) the accretor is a black hole,
and the accreting object spirals within its horizon before filling its
Roche (or more generally tidal) lobe. This occurs if $R_2 < R_{\rm L}$
for $a = \eta G M_1/c^2$. With $R_2 = 10^6R_6$ cm, we find the condition
\begin{equation}
{M_1^3M_2\over M_1 + M_2} > \biggl({7.2R_6\over\eta}\biggr)^3.
\label{sw}
\end{equation}
Only the most massive black holes can swallow neutron stars whole.
This is true for a wide range of neutron--star radii, including
$R_6 = 1$.
Most mergers result in dynamical instability of
the neutron lump. Thus fragmentation and subsequent
coalescence release enough energy to power a gamma--ray burst.

\section{Mergers and kicks}

Simulations of unequal--mass neutron star mergers show that the
mass loss from the system is asymmetric. The escaping material
originates from the lower--mass star and is ejected on a timescale
shorter than the orbital period. This provides a thrust to the merged
object, which is found to have a velocity $V_{\rm kick} \sim
800$ km s$^{-1}$ for the case $M_1 = 0.8$ M$_\odot$ and $M_2 = 0.7$
M$_\odot$ (Rosswog \& Davies, in prep; but see also Rosswog et al,
2000). 

In general we can assume that the ejected material, $M_{\rm lost}$
is ejected at a speed $\propto V_{\rm 2, orb}$ (where $V_{\rm 2,
orb}$ is the orbital velocity of the primary, $M_2$) when the donor
finally gets shredded. Combining expressions for $V_{\rm 2, orb}$ and
the orbital separation $a$ (assuming the donor fills its Roche lobe),
and applying conservation of momentum, we obtain the following
expression for the kick given to the merged object:


\begin{eqnarray}
V_{\rm kick} & \propto & \biggl[{ G M_1^2 \over
(M_1 + M_2) R_2 }{\left( M_2 \over M_1 + M_2 \right)^{1/3}
}\biggr]^{1/2} \nonumber \\
&& \times  { M_{\rm lost} \over M_1 + M_2 - M_{\rm lost} }
\end{eqnarray}

For systems where $M_2$ has been reduced to the mininum mass of $\sim$ 0.2
M$_\odot$, with $M_1 \gg M_2$,
 we see that $V_{\rm kick} \propto M^{-2/3}$ where
$M=M_1+M_2$. We use the result of Rosswog \& Davies (as stated above), 
for $M=1.5$ M$_\odot$, to write

\begin{equation}
V_{\rm kick} = 800 \  { \left( 1.5 M_\odot \over M \right)^{2/3} } \ 
{\rm km~s^{-1}}
\end{equation}

The kick may be slightly lower if the final
shredding occurs before $M_2$ reaches 0.2 M$_\odot$.
It should also be noted that the speed of the compact object at infinity
will be reduced as it is decelerated by the gravitational
force of the ejected material. Likely values of the speed of the merged
object at infinity lie in the range 100 -- 300 km s$^{-1}$, although
the merged object and the ejecta may be bound in some cases.

A remarkably similar kick occurs if both merging objects are black
holes, because of the effect of gravitational radiation reaction on
the final plunge orbit (Bekenstein, 1973; Fitchett, 1983; Fitchett \&
Detweiler, 1984). For rapidly spinning holes, as are likely in core
collapse, the kick velocity may approach 1500 km~s$^{-1}$ (Fitchett,
1983). The basic reason for the similarity is that in both cases the
recoil velocity is of order the primary's centre--of--mass--velocity
immediately before the plunge phase.

\section{Supernova kicks and GRBs}

The torus surrounding the compact object releases its energy into
the region along its rotation axis.
As pointed out by MacFadyen \& Woosley (1999), the production of a GRB
will be inhibited if the volume into which the $\nu - \bar{\nu}$ or
MHD energy is deposited is increased significantly by the motion of
the central object and torus with respect to the surrounding gas. This
motion may be the recoil described in the previous section, or it may
be a kick derived from some other physical mechanism (see e.g. Lai 2001).

A potential GRB will be extinguished if $V_{\rm kick} \go d /
\tau_{\rm er}$, where $d$ is the lengthscale for energy deposition into a
potential fireball, and $\tau_{\rm er}$ is timescale for the energy
to be released from the torus which is set by the viscous
timescale of the torus.  Assuming $ \tau_{\rm er} \sim$ 1 s 
and taking $d = 15 (M_1/M_\odot)$ km (see figures in Fishbone
\& Moncrief, 1976), a GRB will fail if:

\begin{equation}
V_{\rm kick} \ge 15 { \left( M_1 \over M_\odot \right)} \ \ {\rm km~s^{-1}}
\end{equation}



The expression above is plotted in Fig 2 along with the likely
kick received by the central object and torus (as given in equation
5; note here that for the interesting range of values
of $M$, $M_2 \ll M_1$, hence $M_1 \simeq M$). 
This figure suggests that the recoil velocity is likely
to extinguish any potential GRB when the total mass 
$M \lo 12$ M$_\odot$.
In other words GRBs will be
extinguished when $V_{\rm kick}$ exceeds some particular value, 
$\sim 200$ km~s$^{-1}$. The exact limiting mass for a gamma--ray burst
is uncertain, but the important point here is that gamma--ray bursts
will only occur above some limiting mass. 

\begin{figure}
\plotone{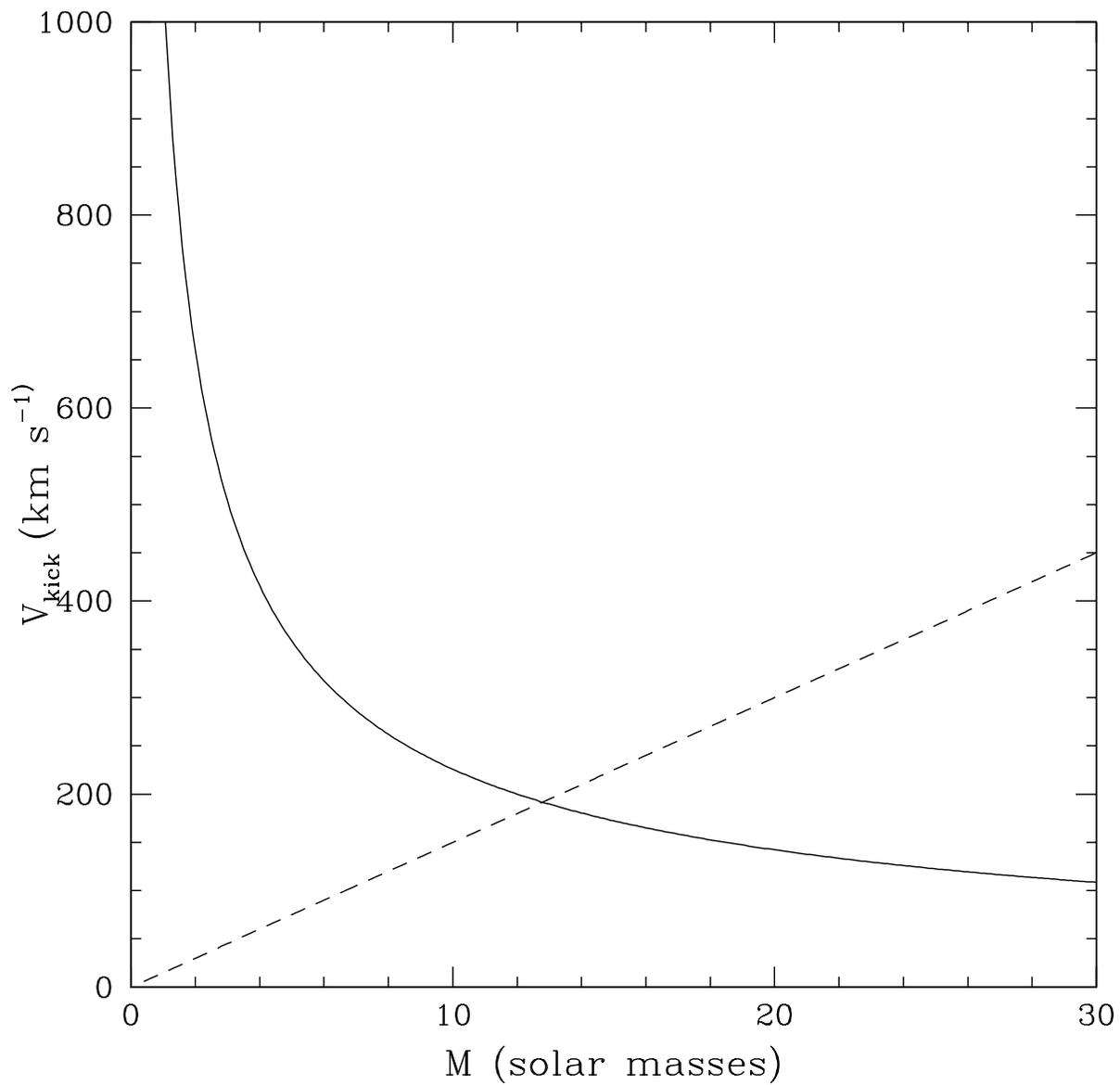}
\caption{The kick received (in km s$^{-1}$) by the central 
object and torus (solid line)
and the kick required to extinguish a GRB (dashed line).
Both are plotted as a function of total mass, $M$ (in solar masses).}
\end{figure}

\section{The SN--GRB delay and gravitational wave emission}

In the picture presented here, core collapse only produces a
gamma--ray burst in a few cases. Even in these cases, the burst may
not follow the collapse immediately, but may be delayed while
gravitational radiation brings the orbiting fragments into
contact. For an initial circular orbit of separation $a_0$, it is easy
to show that this requires a time

\begin{equation}
\tau_{\rm gr} = 
 0.18 \times {(a_0 / 1000 \ {\rm km})^4 \over m_1 m_2 (m_1 + m_2)} \ {\rm hours}.
\end{equation}
where $m_1 = M_1 / $ M$_\odot$ etc. We see 
that a delay of hours is quite possible. Since $\tau_{\rm gr} \propto
a_0^4$, only a small increase in the value of $a_0$ (say to 3000 km),
will produce a large (factor $\sim$ 100) increase in the delay between
formation of the two lumps and their coming into contact. Thus SN--GRB
delays of the order inferred by Reeves et al in GRB 011211 are quite
reasonable in this picture.

An obvious corollary of this is that our picture predicts that
core--collapse supernovae should be strong sources of gravitational
radiation (cf Bonnell \& Pringle, 1995; van Putten, 2001; Fryer, Holz \&
Hughes, 2002). 
A neutron star merger should be detectable by LIGO out to 20 Mpc,
and by LIGO II out to 300 Mpc. The gravitational wave
signal strength for two point masses in circular 
orbits with a separation $a$ is given by $h \propto \Omega^2 \mu a^2$ where 
$\Omega^2 = G (M_1 + M_2) / a^3$ and $\mu = M_1 M_2 /(M_1 + M_2)$.
Hence the detectability of a merger of two compact objects is
a sensitive function of their masses.
As an example, we consider here the  merger of two half neutron stars
(ie $M_1=M_2=0.7$  M$_\odot$.) will be detectable to a distance
of 6 Mpc with LIGO and 100 Mpc with LIGO II. 
We can derive a predicted event rate from an assumed event rate
per galaxy (see Phinney 1991).
Assuming a formation
rate of 10$^{-2}$ yr$^{-1}$ per galaxy, 
LIGO II should see $\sim$ 400 mergers per year.
This is much larger than the number of  neutron--star merger events per year 
LIGO II should detect ($\sim 10$).

\section{Conclusions}

We have suggested by analogy with large--scale simulations of star
formation (see e.g. Bate, Bonnell \& Bromm, 2002, in prep, and
http://www.ukaff.ac.uk/starcluster/) that core collapse of a massive
rotating star may lead to fragmentation of nuclear--density lumps. The
subsequent coalescence of these lumps under gravitational radiation
gives the resulting black hole or neutron star a significant kick
velocity, compatible with those observed in pulsars 
(see for example Arzoumanian, Chernoff, 
\& Cordes, 2002).  A gamma--ray
burst can result only when this kick is small. Thus only a small
fraction of core--collapse supernovae produce gamma--ray bursts.
The most likely candidates are those containing massive
black holes 
($M_1 \go 12$ M$_\odot$)
 which have not formed via
the merger of two lower--mass black holes.
The burst may be delayed significantly (hours -- days)
after the supernova, as suggested by recent observations.

The complexity seen in star formation studies suggests that a large
variety of behaviours is likely in core collapse. A gamma--ray burst
appears to require a rather high degree of symmetry and alignment, and
is therefore a rather unusual outcome. We note that in the case of a
kick driven by mass expulsion in a double neutron--star merger, the
expelled gas may have up to 10 times the energy of the kinetic energy
of the merger product. Given likely initial neutron star kick velocities
of $\sim 1000$ km s$^{-1}$, 
these energies may approach $10^{50}$~erg, and thus have
noticeable effects on the early development of the supernova outburst.

A clear test of our picture will be given by gravitational wave
experiments. An observed chirp signal where the {\em total} mass $\simeq
1.4$ M$_\odot$ would be easily explained in our model but practically
impossible to explain via standard neutron--star mergers (as predicted
from the observed binary pulars [Phinney, 1991]).

\acknowledgements

MBD gratefully acknowledges the support of a URF from the Royal Society.
Theoretical astrophysics at Leicester is supported by a PPARC rolling grant.

\end{document}